\documentclass[a4paper, 12pt]{article}
\usepackage[utf8]{inputenc}
\usepackage[T1]{fontenc}
\usepackage{a4wide, lmodern, amsmath, amsfonts, enumitem, xcolor, graphicx}
\usepackage[colorlinks=true, citecolor=blue, urlcolor=blue, linkcolor=blue, breaklinks=true, pdfpagelabels=false]{hyperref}
\usepackage[numbers, sort&compress]{natbib}

\makeatletter\g@addto@macro\bfseries{\boldmath}\makeatother

\title{LHC EFT WG, Area 1\\Electroweak input parameters}
\author{
Editors:
Ilaria Brivio,
Sally Dawson,
Jorge de Blas,\\
Gauthier Durieux,
Pierre Savard
\\
Contributors:
Ansgar Denner,
Ayres Freitas,
Chris Hays,\\
Ben Pecjak, 
Alessandro Vicini
}

\begin{document}
\makeatletter%
\renewcommand{\@oddhead}{\hfil CERN-LPCC-2021-002}%
\@maketitle%
\makeatother%

\begin{abstract}
Different sets of electroweak input parameters are discussed for SMEFT predictions at the LHC.
The $\{G_\mu,m_Z,m_W\}$ one is presently recommended.
\end{abstract}

A discussion of the advantages and disadvantages of different sets of electroweak input parameters took place in a working-group meeting on \href{https://indico.cern.ch/event/971722}{December 7, 2020} (see also the review on NLO predictions at the \href{https://indico.cern.ch/event/971724/\#8-review-of-nlo-predictions-in}{December 14} meeting).
A preliminary version of this note was presented at the general working-group meeting on \href{https://indico.cern.ch/event/1016713/}{May 3, 2021} and the final recommendation discussed on \href{https://indico.cern.ch/event/1048848/}{June 28}.

Adopting a common set of electroweak parameters for tools making SMEFT predictions for LHC observables would ease comparisons and combinations, even though results obtained with one set can in principle be translated to another.
For the sake of comparison, implementing different choices in tools would still be desirable.

\section{General considerations}

When performing a SMEFT fit to a set of observables, the theoretical expression of these observables is given in terms of SM input parameters and Wilson coefficients for the relevant higher-dimensional operators.
In principle, the input parameters and Wilson coefficients could be fitted to the data simultaneously, and thus they could be treated on equal footing in the fitting process.
However, the practical implementation and performance are improved by extracting input parameters from observables that meet special criteria:
\begin{enumerate}
    \item The ``input observables'' are measured very precisely (such that their experimental uncertainty is subdominant or even negligible in a SMEFT fit).
    \item The extraction of the input parameters from these observables is (to first approximation) independent of any SMEFT effects, but only depends on effects like QED/QCD corrections, hadronization and backgrounds, that can be evaluated reliably within the SM.
    \item When using these input parameters in a SMEFT fit to another set of observables, they do not introduce any dependence on additional unrelated SMEFT operators.
\end{enumerate}
If these criteria are met, the SM input parameters can be determined independently from the actual SMEFT fit, and they can be kept constant during event generation.

\makeatletter%
\renewcommand{\@oddhead}{}%
\makeatother%

The input parameters can be divided into two categories:
\begin{itemize}
    \item On-shell masses that are determined from kinematical features (resonance peaks, thresholds, kinematic endpoints).\footnote{A counterexample would be the determination of the top mass from the total $pp \to t\bar{t}$ cross-section at LHC~\cite{Sirunyan:2019zvx, Aad:2019mkw}, which would be affected by new physics entering via $qqtt$ or $ttg(g)$ operators.}
New physics effects are negligible if the kinematical features are sufficiently sharp.
    \item Gauge couplings:
    \begin{enumerate}
        \item[a)] The electromagnetic coupling $\alpha$ is very precisely determined from very low energy processes~\cite{Aoyama:2019ryr,Parker:2018vye}, where new physics effects decouple and are additionally constrained by QED gauge invariance. For electroweak processes one also needs the running coupling $\alpha(m_Z)=\alpha(0)(1-\Delta\alpha)^{-1}$~\cite{Blondel:2019vdq,Davier:2019can,Keshavarzi:2019abf}, but any new physics effects in $\Delta\alpha$ are also constrained by QED gauge invariance.

For processes involving external on-shell photons, mixed schemes involving also $\alpha(0)$ are advisable (see e.g.\ sec.~5 of ref.~\cite{Denner:2019vbn}).

        \item[b)] The QCD coupling $\alpha_s$ can be determined from many different observables, but the most precise determinations are based on the QCD static potential and lattice QCD (see section on ``Quantum Chromodynamics'' in~\cite{Zyla:2020zbs}), and they are thus protected from new physics by QCD gauge invariance.
        \item[c)] The Fermi constant $G_\mu$ can be extracted from the muon lifetime  \cite{Webber:2010zf}, which depends on QED corrections and possibly also on new physics at scales $\Lambda > m_\mu$.
    \end{enumerate}
\end{itemize}
There is a redundancy in the SM among the parameters $\alpha,G_\mu,m_W,m_Z$, and thus only three of these should be chosen as input parameters. The three most common choices for these three parameters are discussed in the following sections.

\section{\texorpdfstring{$\{\alpha,G_\mu, m_Z\}$}{\{alpha, Gmu, mZ\}}}

This choice is commonly used in electroweak precision studies, since these three parameters are most precisely known among the set of observables considered there. However, $m_W$ must then be computed using the relation 
\begin{align}
\frac{G_\mu}{\sqrt{2}} &= \frac{\pi\alpha}{2s^2c^2m_Z^2}\biggl ( 1 + \Delta r_{\rm SM}
 -2m_W^2 {\frac{c_{\rm BW}}{\Lambda^2}} + \frac{2m_W^2c^2}{e^2}\,
  {\frac{c_{\phi 1}}{\Lambda^2}} \biggr ) - 
  2{\frac{c_{\rm LL}^{(3)\ell}}{\Lambda^2}} + 
  2{\frac{c_{\rm L}^{(3)\ell}}{\Lambda^2}}\,, \\
  c &= m_W/m_Z, \qquad s^2 = 1-c^2.
\end{align}
Here the contributions of dimension-six operators in the HISZ basis~\cite{Hagiwara:1993ck} have been included. This implies that the $W$ mass receives SMEFT corrections that, as a consequence, will enter many LHC observables. Since $m_W$ appears in the propagators and phase-space boundaries, the dependence on these Wilson coefficients will be highly non-linear, which can be impractical for SMEFT fits.
A possible way out is to re-expand the dependence on the Wilson coefficients to linear order. However, even under the assumption that the latter are sufficiently small to allow expansion, this is difficult to achieve in Monte Carlo programs and may also pose theoretical issues.

\section{\texorpdfstring{$\{G_\mu, m_Z, m_W\}$}{\{Gmu, mZ, mW\}}}

Presently, the developers of SMEFT tools for LHC predictions tend to favour the $\{G_\mu$, $m_Z$, $m_W\}$ set.

As it includes both weak gauge boson masses, it has the advantage of reducing SMEFT dependencies in propagators.
(Accounting for the SMEFT dependence of the total widths is an issue to be discussed elsewhere.)
Mass measurements resting on kinematical features also have clear, model independent, interpretations.

However, $G_\mu$ in the SMEFT is modified by higher-dimensional operators. There are a few of these at leading order (with the precise number depending on the basis choice), including the four-fermion operator ${\cal O}_{\rm LL}^{(3)\ell}$, and many more at higher orders. Thus the dependence on these operators must always be taken into account, even for sets of observables that are entirely unrelated to leptonic physics.

As measurements of $m_W$ are actively pursued, predictions including it as an input may need to be recomputed for applications in which small shifts would matter.

\section{\texorpdfstring{$\{\alpha, m_Z, m_W\}$}{\{alpha, mZ, mW\}}}

Using $\alpha$ as input instead of $G_\mu$ has the advantage of avoiding the shifts in electroweak couplings that arise from operator coefficients affecting the muon decay (such as ${\cal O}_{\rm LL}^{(3)\ell}$)~\cite{Gupta:2014rxa}.

It was argued at the \href{https://indico.cern.ch/event/971724/\#8-review-of-nlo-predictions-in}{December 14} meeting that the $\alpha$ scheme leads to slightly worse electroweak convergences in processes involving $g_2=e/\sin\theta_w$ or the Higgs vacuum expectation value, both in the SM and in SMEFT.
Using $G_\mu$ as an input instead of $\alpha$ indeed absorbs $m_t^2/m_W^2$ corrections from the $\rho$ parameter into the lower-order result (see e.g.\ sec.~5 of ref.~\cite{Denner:2019vbn}).
When NLO EW corrections are included in the SM, results obtained in the two schemes are however very close to each other.
Differences with respect to the $\{G_\mu, m_Z, m_W\}$ scheme could be incorporated by resuming contributions from the running of $\alpha$ and including higher-order contributions to the $\rho$ parameter from the standard model.

\section{Conversion and combinations}

Irrespective of the electroweak input set they rely on, electroweak and global fits should include the precise measurements of the other parameters as constraints.
This means that measurements of $m_W$, $\alpha$, or $G_\mu$ need to be included as constraints in fits that respectively employ $\{\alpha,G_\mu,m_Z\}$, $\{G_\mu,m_Z,m_W\}$, or $\{\alpha, m_Z, m_W\}$ as inputs.

If $m_W$ is selected as an input quantity for LHC measurements, a potential issue arises when addressing the combination of LHC and LEP results, as precision calculations of electroweak observables favoured the  $\{\alpha, G_\mu, m_Z\}$ set.
For consistency and up to the desired precision, these would need to be converted. In fact, this conversion is relatively straightforward since many electroweak precision fitters already use $m_W$ as an internal parameter.
Conversions between schemes that involve both $m_Z$ and $m_W$ are also simplified by the fact the coupling dependence is polynomial.

It should be noted that, in principle, the conversion between input schemes concerns two different aspects:
\begin{enumerate}[label=(\roman*)]
\item the predicted value of a given observable \emph{in the SM} is modified. Numerically, this scheme dependence decreases as higher-order perturbative corrections are included.
\label{conversion_sm}

\item the dependence on the Wilson coefficients changes. 
\label{conversion_d6}
\end{enumerate}
Typically, effects due to~\ref{conversion_sm} have a much smaller impact on global fits compared to those from~\ref{conversion_d6}, especially if the observables involved are computed with high accuracy in the SM.

The conversion~\ref{conversion_d6} can be reduced to a simple a Jacobian transformation~\cite{Brivio:2017bnu,Brivio:2020onw}.
Let us consider the translation from the $\mathcal{I}=\{I_1,\dots I_n\}$ input set to the $\mathcal{I}'=\{I'_1,\dots I'_n\}$ one.
Predictions for the $\mathcal{I}$ observables are first obtained in the $\mathcal{I}'$ scheme,
\begin{equation}
\mathcal{I}^\text{SM}(\mathcal{I}') + \delta \mathcal{I}(\mathcal{I}')+ O(C^2/\Lambda^4)\,,
\end{equation}
separating SM contributions and linearised dimension-six SMEFT corrections.
Denoting $\mathcal{O}^\text{SM}+\delta \mathcal{O}$ and $\mathcal{O}^{\prime \text{SM}}+\delta\mathcal{O}'$ the linearised SMEFT predictions for any observable, respectively in the $\mathcal{I}$ and $\mathcal{I}'$ schemes, the translation is achieved by:
\begin{align}
\delta\mathcal{O}'(\mathcal{I}')
	&=
	\delta\mathcal{O}(\mathcal{I}^\text{SM}(\mathcal{I}'))
	+\frac{\partial \mathcal{O}^\text{SM}}{\partial \mathcal{I}}(\mathcal{I}^\text{SM}(\mathcal{I}'))\;\: \delta \mathcal{I}(\mathcal{I}')
\,.
\end{align}
Note that the change in $\delta\mathcal{O}$ due to the shift from the original inputs $\mathcal{I}$ to $\mathcal{I}^\text{SM}(\mathcal{I}')$ is generally expected to be subleading.
When it is neglected, only the functional dependence of $\mathcal{O}^\text{SM}$ on $\mathcal{I}$ is required for the translation, and not that of $\delta\mathcal{O}$.

Alternatively, one can just re-compute the process of interest using a tool set up with a different set of input parameters.
Both methods have advantages and disadvantages. Ultimately, which of the two is preferable depends on the process and desired level of approximation. Some aspects to consider are:
\begin{itemize}
\item The main complication in applying the Jacobian method is that calculating $\partial\mathcal{O}^\text{SM}/\partial \mathcal{I}$ requires (semi-)analytic knowledge of the dependence of the SM prediction on the input quantities (at the perturbative order chosen for the SMEFT calculation).
For differential predictions at the LHC, extracting such a semi-analytical dependence could require a fit of the SM prediction sampled over input parameter values.

\item The conversion to quadratic order in dimension-six operator coefficients has an extra complication in the Jacobian method, because it would also require knowledge about the dependence on the input parameters of the linear SMEFT contributions, $\delta\mathcal{O}$.
It is also needed to account for the dependence of $\delta\mathcal{O}$ itself on the input scheme.

\item  The Jacobian method is convenient for estimating the leading dependence on the Wilson coefficients that directly modify the input quantities, which are typically only a few.
On the other hand, a direct re-calculation is very easy for input schemes that are already implemented in existing tools.
\end{itemize}

The conversion~\ref{conversion_sm} also requires (semi-)analytical knowledge of the functional dependence of the SM prediction on the inputs:
\begin{equation}
\mathcal{O}^{\prime\text{SM}}(\mathcal{I}') = \mathcal{O}^\text{SM}(\mathcal{I}(\mathcal{I}'))\,.
\end{equation}
For instance in ref.~\cite{Brivio:2017bnu, Corbett:2021eux} the SM predictions for EWPD were converted from the $\{\alpha,G_\mu,m_Z\}$ to the $\{G_\mu,m_Z,m_W\}$ scheme employing the semi-analytic formulae provided in ref.~\cite{Freitas:2014hra} (Eq.~(28) and Table~5) and replacing $\Delta\alpha$ by the expression of $\Delta\alpha(m_W)$ obtained solving Eqs.\,(6,7) in ref.~\cite{Awramik:2003rn}.
Because this calculation included higher-order corrections in QCD and QED, the scheme dependence turned out to be numerically very small ($< 1$\textperthousand).

\section{Recommendation}

In view of the discussion above, the LHC EFT WG recommends that the $\{G_\mu,m_Z,m_W\}$ set of electroweak input parameters be presently preferred.
As noted earlier, the implementation in simulation tools of the other two sets discussed here and their comparisons would still be instructive.
This recommendation could be re-assessed in the future, notably with the generalization of electroweak SMEFT corrections.

\bibliographystyle{apsrev4-1_title}
\bibliography{main.bib}

\begin{thebibliography}{17}%
\makeatletter
\providecommand \@ifxundefined [1]{%
 \@ifx{#1\undefined}
}%
\providecommand \@ifnum [1]{%
 \ifnum #1\expandafter \@firstoftwo
 \else \expandafter \@secondoftwo
 \fi
}%
\providecommand \@ifx [1]{%
 \ifx #1\expandafter \@firstoftwo
 \else \expandafter \@secondoftwo
 \fi
}%
\providecommand \natexlab [1]{#1}%
\providecommand \enquote  [1]{``#1''}%
\providecommand \bibnamefont  [1]{#1}%
\providecommand \bibfnamefont [1]{#1}%
\providecommand \citenamefont [1]{#1}%
\providecommand \href@noop [0]{\@secondoftwo}%
\providecommand \href [0]{\begingroup \@sanitize@url \@href}%
\providecommand \@href[1]{\@@startlink{#1}\@@href}%
\providecommand \@@href[1]{\endgroup#1\@@endlink}%
\providecommand \@sanitize@url [0]{\catcode `\\12\catcode `\$12\catcode
  `\&12\catcode `\#12\catcode `\^12\catcode `\_12\catcode `\%12\relax}%
\providecommand \@@startlink[1]{}%
\providecommand \@@endlink[0]{}%
\providecommand \url  [0]{\begingroup\@sanitize@url \@url }%
\providecommand \@url [1]{\endgroup\@href {#1}{\urlprefix }}%
\providecommand \urlprefix  [0]{URL }%
\providecommand \Eprint [0]{\href }%
\providecommand \doibase [0]{http://dx.doi.org/}%
\providecommand \selectlanguage [0]{\@gobble}%
\providecommand \bibinfo [0]{\@secondoftwo}%
\providecommand \bibfield [0]{\@secondoftwo}%
\providecommand \translation [1]{[#1]}%
\providecommand \BibitemOpen [0]{}%
\providecommand \bibitemStop [0]{}%
\providecommand \bibitemNoStop [0]{.\EOS\space}%
\providecommand \EOS [0]{\spacefactor3000\relax}%
\providecommand \BibitemShut  [1]{\csname bibitem#1\endcsname}%
\let\auto@bib@innerbib\@empty
\bibitem [{\citenamefont{Sirunyan} \emph {et\,al.}(2020)}]{Sirunyan:2019zvx}%
  \BibitemOpen
  \bibfield{author}{\bibinfo{author}{\bibfnamefont{A.~M.}
  \bibnamefont{Sirunyan}} \emph {et\,al.} (\bibinfo{collaboration}{CMS}),
  }\bibfield{title}{\emph {\bibinfo{title}{{Measurement of $\mathrm{t\bar t}$
  normalised multi-differential cross sections in pp collisions at $\sqrt s=13$
  TeV, and simultaneous determination of the strong coupling strength, top
  quark pole mass, and parton distribution functions}}}, }\href {\doibase
  10.1140/epjc/s10052-020-7917-7} {\bibfield{journal}{\bibinfo{journal}{Eur.
  Phys. J.
  C}\,}\textbf{\bibinfo{volume}{80}}\,(\bibinfo{year}{2020})\,\bibinfo{pages}{658}},
  \Eprint {http://arxiv.org/abs/1904.05237}{arXiv:1904.05237
  [hep-ex]}\BibitemShut {NoStop}%
\bibitem [{\citenamefont{Aad} \emph {et\,al.}(2019)}]{Aad:2019mkw}%
  \BibitemOpen
  \bibfield{author}{\bibinfo{author}{\bibfnamefont{G.}\,\bibnamefont{Aad}}
  \emph {et\,al.} (\bibinfo{collaboration}{ATLAS}), }\bibfield{title}{\emph
  {\bibinfo{title}{{Measurement of the top-quark mass in $t\bar{t}+1$-jet
  events collected with the ATLAS detector in $pp$ collisions at $\sqrt{s}=8$
  TeV}}}, }\href {\doibase 10.1007/JHEP11(2019)150}
  {\bibfield{journal}{\bibinfo{journal}{JHEP}\,}\textbf{\bibinfo{volume}{11}}\,(\bibinfo{year}{2019})\,\bibinfo{pages}{150}},
  \Eprint {http://arxiv.org/abs/1905.02302}{arXiv:1905.02302
  [hep-ex]}\BibitemShut {NoStop}%
\bibitem [{\citenamefont{Aoyama} \emph {et\,al.}(2019)\citenamefont{Aoyama},
  \citenamefont{Kinoshita}, and \citenamefont{Nio}}]{Aoyama:2019ryr}%
  \BibitemOpen
  \bibfield{author}{\bibinfo{author}{\bibfnamefont{T.}\,\bibnamefont{Aoyama}},
  \bibinfo{author}{\bibfnamefont{T.}\,\bibnamefont{Kinoshita}},  and
  \bibinfo{author}{\bibfnamefont{M.}\,\bibnamefont{Nio}},
  }\bibfield{title}{\emph {\bibinfo{title}{{Theory of the Anomalous Magnetic
  Moment of the Electron}}}, }\href {\doibase 10.3390/atoms7010028}
  {\bibfield{journal}{\bibinfo{journal}{Atoms}\,}\textbf{\bibinfo{volume}{7}}\,(\bibinfo{year}{2019})\,\bibinfo{pages}{28}}\BibitemShut
  {NoStop}%
\bibitem [{\citenamefont{Parker} \emph {et\,al.}(2018)\citenamefont{Parker},
  \citenamefont{Yu}, \citenamefont{Zhong}, \citenamefont{Estey}, and
  \citenamefont{M\"uller}}]{Parker:2018vye}%
  \BibitemOpen
  \bibfield{author}{\bibinfo{author}{\bibfnamefont{R.~H.}
  \bibnamefont{Parker}}, \bibinfo{author}{\bibfnamefont{C.}\,\bibnamefont{Yu}},
  \bibinfo{author}{\bibfnamefont{W.}\,\bibnamefont{Zhong}},
  \bibinfo{author}{\bibfnamefont{B.}\,\bibnamefont{Estey}},  and
  \bibinfo{author}{\bibfnamefont{H.}\,\bibnamefont{M\"uller}},
  }\bibfield{title}{\emph {\bibinfo{title}{{Measurement of the fine-structure
  constant as a test of the Standard Model}}}, }\href {\doibase
  10.1126/science.aap7706}
  {\bibfield{journal}{\bibinfo{journal}{Science}\,}\textbf{\bibinfo{volume}{360}}\,(\bibinfo{year}{2018})\,\bibinfo{pages}{191}},
  \Eprint {http://arxiv.org/abs/1812.04130}{arXiv:1812.04130
  [physics.atom-ph]}\BibitemShut {NoStop}%
\bibitem [{\citenamefont{Blondel} \emph {et\,al.}(2019)\citenamefont{Blondel},
  \citenamefont{Gluza}, \citenamefont{Jadach}, \citenamefont{Janot}, and
  \citenamefont{Riemann}}]{Blondel:2019vdq}%
  \BibitemOpen
  \bibinfo{editor}{\bibfnamefont{A.}\,\bibnamefont{Blondel}},
  \bibinfo{editor}{\bibfnamefont{J.}\,\bibnamefont{Gluza}},
  \bibinfo{editor}{\bibfnamefont{S.}\,\bibnamefont{Jadach}},
  \bibinfo{editor}{\bibfnamefont{P.}\,\bibnamefont{Janot}},  and
  \bibinfo{editor}{\bibfnamefont{T.}\,\bibnamefont{Riemann}}, eds., \href
  {\doibase 10.23731/CYRM-2020-003} {\emph {\bibinfo{title}{{Theory for the
  FCC-ee}: {Report on the 11th FCC-ee Workshop Theory and Experiments}}}},
  \bibinfo{series}{CERN Yellow Reports: Monographs}, Vol.
  \bibinfo{volume}{3/2020}\,(\bibinfo{publisher}{CERN},
  \bibinfo{address}{Geneva}, \bibinfo{year}{2019})\,\Eprint
  {http://arxiv.org/abs/1905.05078}{arXiv:1905.05078 [hep-ph]}\BibitemShut
  {NoStop}%
\bibitem [{\citenamefont{Davier} \emph {et\,al.}(2020)\citenamefont{Davier},
  \citenamefont{Hoecker}, \citenamefont{Malaescu}, and
  \citenamefont{Zhang}}]{Davier:2019can}%
  \BibitemOpen
  \bibfield{author}{\bibinfo{author}{\bibfnamefont{M.}\,\bibnamefont{Davier}},
  \bibinfo{author}{\bibfnamefont{A.}\,\bibnamefont{Hoecker}},
  \bibinfo{author}{\bibfnamefont{B.}\,\bibnamefont{Malaescu}},  and
  \bibinfo{author}{\bibfnamefont{Z.}\,\bibnamefont{Zhang}},
  }\bibfield{title}{\emph {\bibinfo{title}{{A new evaluation of the hadronic
  vacuum polarisation contributions to the muon anomalous magnetic moment and
  to $\mathbf{\boldsymbol\alpha(m_Z^2)}$}}}, }\href {\doibase
  10.1140/epjc/s10052-020-7792-2} {\bibfield{journal}{\bibinfo{journal}{Eur.
  Phys. J.
  C}\,}\textbf{\bibinfo{volume}{80}}\,(\bibinfo{year}{2020})\,\bibinfo{pages}{241}},
  \bibinfo{note}{[Erratum: Eur.Phys.J.C 80, 410 (2020)]}, \Eprint
  {http://arxiv.org/abs/1908.00921}{arXiv:1908.00921 [hep-ph]}\BibitemShut
  {NoStop}%
\bibitem [{\citenamefont{Keshavarzi} \emph
  {et\,al.}(2020)\citenamefont{Keshavarzi}, \citenamefont{Nomura}, and
  \citenamefont{Teubner}}]{Keshavarzi:2019abf}%
  \BibitemOpen
  \bibfield{author}{\bibinfo{author}{\bibfnamefont{A.}\,\bibnamefont{Keshavarzi}},
  \bibinfo{author}{\bibfnamefont{D.}\,\bibnamefont{Nomura}},  and
  \bibinfo{author}{\bibfnamefont{T.}\,\bibnamefont{Teubner}},
  }\bibfield{title}{\emph {\bibinfo{title}{{$g-2$ of charged leptons, $\alpha
  (M^2_Z)$ , and the hyperfine splitting of muonium}}}, }\href {\doibase
  10.1103/PhysRevD.101.014029} {\bibfield{journal}{\bibinfo{journal}{Phys. Rev.
  D}\,}\textbf{\bibinfo{volume}{101}}\,(\bibinfo{year}{2020})\,\bibinfo{pages}{014029}},
  \Eprint {http://arxiv.org/abs/1911.00367}{arXiv:1911.00367
  [hep-ph]}\BibitemShut {NoStop}%
\bibitem [{\citenamefont{Denner} and
  \citenamefont{Dittmaier}(2020)}]{Denner:2019vbn}%
  \BibitemOpen
  \bibfield{author}{\bibinfo{author}{\bibfnamefont{A.}\,\bibnamefont{Denner}}
  and \bibinfo{author}{\bibfnamefont{S.}\,\bibnamefont{Dittmaier}},
  }\bibfield{title}{\emph {\bibinfo{title}{{Electroweak Radiative Corrections
  for Collider Physics}}}, }\href {\doibase 10.1016/j.physrep.2020.04.001}
  {\bibfield{journal}{\bibinfo{journal}{Phys.
  Rept.}\,}\textbf{\bibinfo{volume}{864}}\,(\bibinfo{year}{2020})\,\bibinfo{pages}{1}},
  \Eprint {http://arxiv.org/abs/1912.06823}{arXiv:1912.06823
  [hep-ph]}\BibitemShut {NoStop}%
\bibitem [{\citenamefont{Zyla} \emph {et\,al.}(2020)}]{Zyla:2020zbs}%
  \BibitemOpen
  \bibfield{author}{\bibinfo{author}{\bibfnamefont{P.~A.} \bibnamefont{Zyla}}
  \emph {et\,al.} (\bibinfo{collaboration}{Particle Data Group}),
  }\bibfield{title}{\emph {\bibinfo{title}{{Review of Particle Physics}}},
  }\href {\doibase 10.1093/ptep/ptaa104}
  {\bibfield{journal}{\bibinfo{journal}{PTEP}\,}\textbf{\bibinfo{volume}{2020}}\,(\bibinfo{year}{2020})\,\bibinfo{pages}{083C01}}\BibitemShut
  {NoStop}%
\bibitem [{\citenamefont{Webber} \emph {et\,al.}(2011)}]{Webber:2010zf}%
  \BibitemOpen
  \bibfield{author}{\bibinfo{author}{\bibfnamefont{D.~M.} \bibnamefont{Webber}}
  \emph {et\,al.} (\bibinfo{collaboration}{MuLan}), }\bibfield{title}{\emph
  {\bibinfo{title}{{Measurement of the Positive Muon Lifetime and Determination
  of the Fermi Constant to Part-per-Million Precision}}}, }\href {\doibase
  10.1103/PhysRevLett.106.079901} {\bibfield{journal}{\bibinfo{journal}{Phys.
  Rev.
  Lett.}\,}\textbf{\bibinfo{volume}{106}}\,(\bibinfo{year}{2011})\,\bibinfo{pages}{041803}},
  \Eprint {http://arxiv.org/abs/1010.0991}{arXiv:1010.0991
  [hep-ex]}\BibitemShut {NoStop}%
\bibitem [{\citenamefont{Hagiwara} \emph
  {et\,al.}(1993)\citenamefont{Hagiwara}, \citenamefont{Ishihara},
  \citenamefont{Szalapski}, and \citenamefont{Zeppenfeld}}]{Hagiwara:1993ck}%
  \BibitemOpen
  \bibfield{author}{\bibinfo{author}{\bibfnamefont{K.}\,\bibnamefont{Hagiwara}},
  \bibinfo{author}{\bibfnamefont{S.}\,\bibnamefont{Ishihara}},
  \bibinfo{author}{\bibfnamefont{R.}\,\bibnamefont{Szalapski}},  and
  \bibinfo{author}{\bibfnamefont{D.}\,\bibnamefont{Zeppenfeld}},
  }\bibfield{title}{\emph {\bibinfo{title}{{Low-energy effects of new
  interactions in the electroweak boson sector}}}, }\href {\doibase
  10.1103/PhysRevD.48.2182} {\bibfield{journal}{\bibinfo{journal}{Phys. Rev.
  D}\,}\textbf{\bibinfo{volume}{48}}\,(\bibinfo{year}{1993})\,\bibinfo{pages}{2182}}\BibitemShut
  {NoStop}%
\bibitem [{\citenamefont{Gupta} \emph {et\,al.}(2015)\citenamefont{Gupta},
  \citenamefont{Pomarol}, and \citenamefont{Riva}}]{Gupta:2014rxa}%
  \BibitemOpen
  \bibfield{author}{\bibinfo{author}{\bibfnamefont{R.~S.} \bibnamefont{Gupta}},
  \bibinfo{author}{\bibfnamefont{A.}\,\bibnamefont{Pomarol}},  and
  \bibinfo{author}{\bibfnamefont{F.}\,\bibnamefont{Riva}},
  }\bibfield{title}{\emph {\bibinfo{title}{{BSM Primary Effects}}}, }\href
  {\doibase 10.1103/PhysRevD.91.035001}
  {\bibfield{journal}{\bibinfo{journal}{Phys. Rev.
  D}\,}\textbf{\bibinfo{volume}{91}}\,(\bibinfo{year}{2015})\,\bibinfo{pages}{035001}},
  \Eprint {http://arxiv.org/abs/1405.0181}{arXiv:1405.0181
  [hep-ph]}\BibitemShut {NoStop}%
\bibitem [{\citenamefont{Brivio} and
  \citenamefont{Trott}(2017)}]{Brivio:2017bnu}%
  \BibitemOpen
  \bibfield{author}{\bibinfo{author}{\bibfnamefont{I.}\,\bibnamefont{Brivio}}
  and \bibinfo{author}{\bibfnamefont{M.}\,\bibnamefont{Trott}},
  }\bibfield{title}{\emph {\bibinfo{title}{{Scheming in the SMEFT... and a
  reparameterization invariance!}}}, }\href {\doibase 10.1007/JHEP07(2017)148}
  {\bibfield{journal}{\bibinfo{journal}{JHEP}\,}\textbf{\bibinfo{volume}{07}}\,(\bibinfo{year}{2017})\,\bibinfo{pages}{148}},
  \bibinfo{note}{[Addendum: JHEP 05, 136 (2018)]}, \Eprint
  {http://arxiv.org/abs/1701.06424}{arXiv:1701.06424 [hep-ph]}\BibitemShut
  {NoStop}%
\bibitem [{\citenamefont{Brivio}(2021)}]{Brivio:2020onw}%
  \BibitemOpen
  \bibfield{author}{\bibinfo{author}{\bibfnamefont{I.}\,\bibnamefont{Brivio}},
  }\bibfield{title}{\emph {\bibinfo{title}{{SMEFTsim 3.0 \textemdash{} a
  practical guide}}}, }\href {\doibase 10.1007/JHEP04(2021)073}
  {\bibfield{journal}{\bibinfo{journal}{JHEP}\,}\textbf{\bibinfo{volume}{04}}\,(\bibinfo{year}{2021})\,\bibinfo{pages}{073}},
  \Eprint {http://arxiv.org/abs/2012.11343}{arXiv:2012.11343
  [hep-ph]}\BibitemShut {NoStop}%
\bibitem [{\citenamefont{Corbett} \emph {et\,al.}(2021)\citenamefont{Corbett},
  \citenamefont{Helset}, \citenamefont{Martin}, and
  \citenamefont{Trott}}]{Corbett:2021eux}%
  \BibitemOpen
  \bibfield{author}{\bibinfo{author}{\bibfnamefont{T.}\,\bibnamefont{Corbett}},
  \bibinfo{author}{\bibfnamefont{A.}\,\bibnamefont{Helset}},
  \bibinfo{author}{\bibfnamefont{A.}\,\bibnamefont{Martin}},  and
  \bibinfo{author}{\bibfnamefont{M.}\,\bibnamefont{Trott}},
  }\bibfield{title}{\emph {\bibinfo{title}{{EWPD in the SMEFT to dimension
  eight}}}, }\href {\doibase 10.1007/JHEP06(2021)076}
  {\bibfield{journal}{\bibinfo{journal}{JHEP}\,}\textbf{\bibinfo{volume}{06}}\,(\bibinfo{year}{2021})\,\bibinfo{pages}{076}},
  \Eprint {http://arxiv.org/abs/2102.02819}{arXiv:2102.02819
  [hep-ph]}\BibitemShut {NoStop}%
\bibitem [{\citenamefont{Freitas}(2014)}]{Freitas:2014hra}%
  \BibitemOpen
  \bibfield{author}{\bibinfo{author}{\bibfnamefont{A.}\,\bibnamefont{Freitas}},
  }\bibfield{title}{\emph {\bibinfo{title}{{Higher-order electroweak
  corrections to the partial widths and branching ratios of the Z boson}}},
  }\href {\doibase 10.1007/JHEP04(2014)070}
  {\bibfield{journal}{\bibinfo{journal}{JHEP}\,}\textbf{\bibinfo{volume}{04}}\,(\bibinfo{year}{2014})\,\bibinfo{pages}{070}},
  \Eprint {http://arxiv.org/abs/1401.2447}{arXiv:1401.2447
  [hep-ph]}\BibitemShut {NoStop}%
\bibitem [{\citenamefont{Awramik} \emph {et\,al.}(2004)\citenamefont{Awramik},
  \citenamefont{Czakon}, \citenamefont{Freitas}, and
  \citenamefont{Weiglein}}]{Awramik:2003rn}%
  \BibitemOpen
  \bibfield{author}{\bibinfo{author}{\bibfnamefont{M.}\,\bibnamefont{Awramik}},
  \bibinfo{author}{\bibfnamefont{M.}\,\bibnamefont{Czakon}},
  \bibinfo{author}{\bibfnamefont{A.}\,\bibnamefont{Freitas}},  and
  \bibinfo{author}{\bibfnamefont{G.}\,\bibnamefont{Weiglein}},
  }\bibfield{title}{\emph {\bibinfo{title}{{Precise prediction for the W boson
  mass in the standard model}}}, }\href {\doibase 10.1103/PhysRevD.69.053006}
  {\bibfield{journal}{\bibinfo{journal}{Phys. Rev.
  D}\,}\textbf{\bibinfo{volume}{69}}\,(\bibinfo{year}{2004})\,\bibinfo{pages}{053006}},
  \Eprint
  {http://arxiv.org/abs/hep-ph/0311148}{arXiv:hep-ph/0311148}\BibitemShut
  {NoStop}%
\end{thebibliography}%

\end{document}